\documentclass[11pt,preprint]{aastex}
\usepackage{color}
\usepackage{epsfig}
\def\lsim{\raise0.3ex\hbox{$<$}\kern-0.75em{\lower0.65ex\hbox{$\sim$}}}
\def\gsim{\raise0.3ex\hbox{$>$}\kern-0.75em{\lower0.65ex\hbox{$\sim$}}}

\begin{document}

\title{Confronting Outflow-Regulated Cluster Formation Model with
Observations}
\author{Fumitaka Nakamura}
\affil{National Astronomical Observatory, Mitaka, Tokyo 181-8588, 
Japan}
\affil{Nobeyama Radio Observatory, Minamimaki, Minamisaku, 
Nagano 384-1805, Japan; fumitaka.nakamura@nao.ac.jp}
\and 
\author{Zhi-Yun Li}
\affil{Department of Astronomy, University of Virginia,
P. O. Box 400325, Charlottesville, VA 22904; zl4h@virginia.edu}

\begin{abstract}
Protostellar outflows have been shown theoretically to be capable of
 maintaining  supersonic turbulence in cluster-forming clumps and 
keeping the star formation rate per free-fall time as low as a few
percent.  
We aim to test two basic predictions of this
outflow-regulated cluster formation  model, namely 
%
(1) the clump should be close to virial equilibrium 
and (2) the turbulence dissipation rate should be balanced by the
 outflow momentum injection rate, using recent outflow surveys toward 
8 nearby cluster-forming clumps  (B59, L1551, L1641N, Serpens Main 
Cloud, Serpens  South, $\rho$ Oph, IC 348, and NGC 1333). 
We find, for almost all sources, that the clumps are close to virial 
equilibrium and the outflow momentum injection rate exceeds 
the turbulence momentum dissipation rate. In addition, the outflow 
kinetic energy is significantly smaller than the clump
 gravitational energy for intermediate and massive clumps 
with $M_{\rm cl} \gtrsim {\rm a \ few} \times 10^2 M_\odot$, 
suggesting that the outflow feedback is not enough to disperse 
the clump as a whole. The number of observed protostars also 
indicates that the star formation rate per free-fall time is as 
small as a few percent for all clumps.
These observationally-based results strengthen the case for outflow-regulated cluster formation. 
\end{abstract}

\section{Introduction}
\label{sec:intro}

There have been many efforts aiming at understanding the formation
process of 
star clusters, the birthplace of the majority of stars 
\citep[e.g.,][and reference therein]{allen07,mckee07}. 
Recent theoretical studies suggest that stellar feedback 
such as protostellar outflows and stellar radiation 
is a key to understanding the process of star formation in clustered
environments ({\it outflow}: Li \& Nakamura 2006; Matzner 2007;
Nakamura \& Li 2007; Cunninghan et al. 2009; Carroll et al. 2009, 
{\it radiation}: Fall et al. 2010; Peters et al. 2010; Dib 2011; 
Colin et al. 2013, {\it both}: Murray et al. 2010; Hansen et al. 2012).
However, the exact role of stellar feedback in clustered 
star formation remains controversial. 

Two main scenarios have been proposed for the role
of stellar feedback in clustered star formation.
In the first scenario, stellar feedback is envisioned to
destroy the dense cluster-forming clump as a whole, which terminates  
further star formation.
In this case, star formation should be 
rapid and brief \citep{elmegreen07,hartmann07}.
Because supersonic turbulence decays very quickly, 
star formation needs to be terminated within a couple of
turbulent crossing times, to achieve low star 
formation efficiencies (SFEs) that are often observed in 
nearby cluster-forming regions. 
The magnetic field is also considered to play only a minor 
or negligible role in this scenario, where 
the global gravitational collapse leads to rapid star formation.
Hereafter, we refer to this scenario as the rapid star 
formation, where the primary role of stellar feedback, 
particularly radiation feedback from massive stars,
is to terminate star formation quickly.

In contrast, in the second scenario, the stellar feedback is envisioned 
to play the role of maintaining the internal turbulent motions of the 
clumps. Here, the star formation should be slow and can last for several 
free-fall times or longer \citep{tan06,li06,nakamura07}, 
as the dissipated turbulence is replenished by stellar feedback 
\citep{carroll10,hansen12,wang10}.  The magnetic field is also considered to play an 
important role in slowing down the global collapse and further star formation
\citep{nakamura07,tilley07,price08,wang10}.
The parent clump is kept close to a quasi-virial equilibrium by the 
combination of stellar feedback and magnetic field.
Hereafter, we refer to this scenario as the slow star formation.

Since the stellar feedback plays a very different role for these two
scenarios, clarifying its role in clustered star formation
is crucial to constrain how clustered star formation proceeds.
In the present paper, we focus on the role of protostellar outflow 
feedback among various stellar feedback mechanisms because 
the outflow feedback is likely to be a leading mechanism 
for regulating star formation in nearby cluster-forming regions 
where no UV light-emitting massive stars are formed.
Even for high-mass-star-forming regions, the outflow feedback 
is expected to play a key role in regulating star formation at least 
in the early and intermediate stages before massive stars form
\citep[e.g.,][]{li06,wang10}.
We refer to this slow star formation scenario as the outflow-regulated
cluster formation. 

A number of previous studies have attempted to address how the kinetic
energies of the observed outflows influence the ambient gas
for individual cluster-forming clumps 
\citep[e.g.,][]{hatchell07,stanke07,swift08,maury09,arce10,curtis10,
nakamura11a,nakamura11b,narayanan12}.
The main conclusion of these studies is that the energy injection rate
by molecular outflows is generally larger than the turbulence enegy
dissipation rate, and thus 
the outflow feedback has enough energy to sustain the turbulent motions. 
However, the outflow feedback is a momentum-driven feedback because 
radiative energy loss is efficient in the clumps \citep{fall10,krumholz14}.
Here, we compile the outflow data of several cluster-forming clumps,
and verify the role of outflow feedback in cluster-forming clumps by
using the momentum dissipation and injection rates, in addition to the
energy dissipation and injection rates.

To facilitate comparison with observations, we will 
first construct an analytic version of the outflow-regulated cluster
formation model and explore its observational consequences in 
Section \ref{sec:model}. We compare the model 
with the results of molecular outflow surveys
toward nearby cluster-forming regions in Section \ref{sec:obs}. 
Finally, we summarize the main conclusion 
in Section \ref{sec:conclusion}.

\section{Analytic Model of Outflow-Regulated Cluster Formation}
\label{sec:model}

Recent numerical simulations have demonstrated that 
protostellar outflows can indeed inject 
turbulent motions into cluster-forming clumps 
\citep{li06,nakamura07,carroll09,carroll10,wang10,hansen12}.
A common drawback of this type of simulations is the use of periodic boundary conditions, which prevent the outflows from leaving the simulation box, leading to an overestimate of the efficiency of outflow feedback.
However, \citet{wang10} reduced the speed of the outflows before they leave  
the computation box and reached the same conclusion. 
Thus, the periodic boundary condition does not change the main conclusion. 
\citet{nakamura11b} analytically estimated the star formation rate per 
free-fall time of a parent clump on the basis of 
the outflow-regulated cluster formation scenario
\citep[see also][]{matzner07}.
They assumed (1) that the turbulence momentum dissipation rate
balances the outflow momentum injection 
and (2) that the clump is kept close to a virial equilibrium with 
the internal turbulent speed equal to the virial speed.
The numerical simulations of cluster formation 
have shown that the above two conditions are 
reasonably achieved \citep{li06,nakamura07,wang10,hansen12}.
Here, assuming the above two conditions, we
derive several physical quantities that can be 
compared directly with the observations.

\subsection{Turbulence Momentum Dissipation Rate}

Consider a clump with mass $M_{\rm cl}$ and radius $R_{\rm cl}$.
The mean density and column density can be calculated, respectively, as
\begin{equation}
\rho = {M_{\rm cl} \over 4\pi R_{\rm cl}^3/3}
= 6.5 \times 10^{-20} \ {\rm g \ cm}^{-3}
\left({M_{\rm cl} \over 500 \ M_\odot}\right)
\left({R_{\rm cl} \over 0.5 \ {\rm pc}}\right)^{-3} \ ,
\end{equation}
and 
\begin{equation}
\Sigma={M_{\rm cl} \over \pi R_{\rm cl}^2}
=  0.13 \ {\rm g \ cm}^{-2}
\left({M_{\rm cl} \over 500 \ M_\odot}\right)
\left({R_{\rm cl} \over 0.5 \ {\rm pc}}\right)^{-2} \ .
\end{equation}

In the outflow-regulated cluster formation model, the outflow 
feedback replenishes the internal supersonic turbulence
and the clump is kept close to a quasi-virial equilibrium 
\citep{li06,nakamura07,nakamura11c}.
In that case, the momentum dissipation rate of the internal turbulent motion
in the clump, $dP_{\rm turb}/dt$, should balance the outflow momentum
injection rate, $dP_{\rm out}/dt$, as 
\begin{equation}
 {dP_{\rm turb} \over dt} + {dP_{\rm out} \over dt} = 0  \  ,
\label{eq:turbulence}
\end{equation}
where 
the momentum dissipation rate of the internal turbulent motion 
is defined as 
\begin{eqnarray}
 {dP_{\rm turb} \over dt} &=& -0.21 {M_{\rm cl} \sigma_{\rm 3D}
\over t_{\rm diss}} \ , \nonumber \\
&\simeq&  -6.4 \times 10^{-4} M_\odot \ {\rm km \ s}^{-1} \ {\rm yr}^{-1} 
\left({M_{\rm cl} \over 500 \ M_\odot}\right)
\left({R_{\rm cl} \over 0.5 \  {\rm pc}}\right)^{-1}
\left({\sigma_{\rm 1D} \over {\rm km \ s}^{-1}}\right)^2
\ ,
\label{eq:turbdiss} 
\end{eqnarray}
where $\sigma_{\rm 3D}$ is 
the three-dimensional velocity dispersion 
and $\sigma_{\rm 1D}$ (= $3^{-1/2} \sigma_{\rm 3D}$) 
is one-dimensional velocity dispersion, which is 
smaller than the Full-Width-at-Half-Maximum (FWHM) line width 
by a factor of $\sqrt{8 \ln 2}$. 
The momentum dissipation time $t_{\rm diss}$ is given by
\begin{eqnarray}
 t_{\rm diss} &=& {R_{\rm cl} \over \sigma_{\rm 3D}}  \nonumber \\
&\simeq& 0.28 {\rm Myr} \left({R_{\rm cl} \over 0.5 \ {\rm pc}}\right)
\left({\sigma_{\rm 1D} \over {\rm km \ s}^{-1}}\right)^{-1}    \  .
\end{eqnarray}
Here, Equation (\ref{eq:turbdiss}) is derived from 
Equation (8) of \citet{maclow99},
by assuming that the clump mass 
is constant and that the turbulence driving length scale is equal 
to the clump diameter of $2R_{\rm cl}$. 
In the outflow-regulated cluster formation model, 
the internal three-dimensional velocity dispersion should be equal to 
the virial speed as 
\begin{eqnarray}
\sigma_{\rm 3D} &=& \sqrt{\frac{3}{5}af_{\rm B}\frac{GM_{\rm cl}}{R_{\rm cl}}}
\nonumber \ ,  \\
&\simeq& 1.5 \ {\rm km \ s}^{-1}
 \left({a \over 5/3}\right)^{1/2} \left({f_{\rm B} \over 0.5}\right)^{1/2}
\left({M_{\rm cl} \over 500 \ M_\odot}\right)^{1/2}
\left({R_{\rm cl} \over 0.5 \ {\rm pc}}\right)^{-1/2}
\label{eq:virial velocity}
\end{eqnarray}
where the order-of-unity dimensionless parameter, $a$, accounts for 
the effects of density distribution in the clump gravitational energy.
For a uniform sphere and a centrally-condensed sphere with $\rho \propto r^{-2}$, 
$a$ is equal to 1 and 5/3, respectively.
Here, we adopt $a=5/3$ because the cluster-forming clumps 
tend to be centrally-condensed. 
We take into account the magnetic support by multiplying 
the virial speed by a factor $f_{\rm B}$, where
$0 \lesssim f_{\rm B} \lesssim 1$.

\subsection{Momentum Injection Rate due to Protostellar Outflow Feedback}

The outflow momentum injection rate is defined as
\begin{equation}
\frac{dP_{\rm out}}{dt} = f_{\rm out} \epsilon_{\rm SFR} f_W P_* \ ,
\end{equation}
where $\epsilon _{\rm SFR}$ is the star formation rate, 
and $P_*$ the momentum per one solar mass of star formed
\citep{matzner00}, $f_{\rm w}$ the fraction of the outflow
that contributes to the generation of molecular outflows, and 
$f_{\rm out}$ the fraction of the molecular outflow momentum 
that is converted into the clump internal turbulence momentum.
The value $f_{\rm out}$ is uncertain, but is expected to be in 
the range of 0.1 to 1, following the numerical simulations.
According to Figure 2 of \citet{nakamura07}, the clump with 
about $10^3 M_\odot$ has reached a quasi-virial equilibrium at the 
velocity dispersion of about 5 $c_s$ in a few clump free-fall times 
$t_{\rm ff}$, where $c_s$ is the isothermal sound speed.
The total amount of the specific momentum injected into the clump almost linearly increases
with time and reaches about 10 $c_s$ in $3 t_{\rm ff}$, where the free-fall time
is estimated by taking into account the fact that the clump mean density
has increased by a factor of a few in a quasi-virial equilibrium state.
The outflow momentum injection rate is estimated to be 
$f_{\rm out}M_{\rm cl} \times 10 c_s / 3 t_{\rm ff}$, where we made the approximation that the clump
mass is constant because the total mass of stars formed is only a few
percent the clump mass. On the other hand, the turbulence dissipation
rate is roughly estimated to be 
$0.21 M_{\rm cl} \times 5 c_s / t_{\rm diss}$ and the dissipation time $t_{\rm diss}$ is
comparable to $t_{\rm ff}$.
Balancing the momentum injection rate agaist the turbulence 
dissipation rate yields $f_{\rm out} \sim 0.3$.

The actual value of $f_{\rm out}$ may depend on the mass and size of the clump. 
For less massive, small clumps, $f_{\rm out}$ may be smaller because the
outflow lobes are easier to break out of the clump.
The typical outflow speed $V_W$ is about $10^2$ km s$^{-1}$ 
\citep[e.g.,][]{matzner00}, which is assumed to be constant, independent
of the stellar mass.
Replacing $\epsilon_{\rm SFR}$ by the star formation rate per free-fall
time SFR$_{\rm ff}$, Equation (\ref{eq:outflow}) is rewritten as
\begin{eqnarray}
 \frac{dP_{\rm out}}{dt} &=& f_{\rm out} 
{\rm SFR}_{\rm ff} {M_{\rm cl} \over t_{\rm ff}}
 f_{\rm W} V_{\rm W}   \nonumber \\
&\simeq& 2.3 \times 10^{-4} M_\odot {\rm km \ s}^{-1} \ {\rm yr}^{-1}
\left(f_{\rm out} \over 0.3 \right)
\left({{\rm SFR}_{\rm ff} \over 0.01}\right)
\left({f_{\rm W} \over 0.4}\right)
\left({V_{\rm W} \over 100 \  {\rm km \ s}^{-1}}\right)
\left({M_{\rm cl} \over 500 \ M_\odot}\right)^{3/2}
\left({R_{\rm cl} \over 0.5 \ {\rm pc}}\right)^{-3/2} \ ,
\label{eq:outflow}
\end{eqnarray}
where the free-fall time is defined as 
\begin{eqnarray}
t_{\rm ff} &=& \sqrt{3\pi \over 32G\rho}     \nonumber \\
&\simeq& 0.26 \ {\rm Myr} \left({M_{\rm cl} \over 500 \ M_\odot}\right)^{-1/2}
\left({R_{\rm cl} \over 0.5 \ {\rm pc}}\right)^{3/2} \ .
\end{eqnarray}

\subsection{Star Formation Rate Per Free-Fall Time}

Using Equations (\ref{eq:turbulence}), (\ref{eq:turbdiss}),
(\ref{eq:virial velocity}), and (\ref{eq:outflow}), 
the star formation rate per free-fall time expected from 
the outflow-regulated cluster
formation model is given by
\begin{eqnarray}
{\rm SFR}_{\rm ff} &\simeq& 
0.13 a f_{\rm B} f_{\rm out} f_{\rm w}^{-1}V_{\rm w}^{-1} 
\frac{GM_{\rm cl}}{R_{\rm cl}^2} t_{\rm ff}  \nonumber \\
&=& 0.02 
\left(\frac{f_{\rm B}}{0.5}\right) 
\left(f_{\rm out} \over 0.3 \right)^{-1}
\left(\frac{f_{\rm w}}{0.4}\right)^{-1}
\left(\frac{V_{\rm w}}{10^2 \ {\rm km \ s^{-1}}}\right)^{-1}
 \left(\frac{M_{\rm cl}}{500 \ M_\odot}\right)^{1/2}
\left(\frac{R_{\rm cl}}{0.5 \ {\rm pc}}\right)^{-1/2} \ .
\label{eq:sfr_ff}
\end{eqnarray}

For the fiducial numbers of the physical quantities of nearby
cluster-forming clumps, the star formation rate per free-fall time 
is estimated to be a few percent, implying that the 
outflow-regulated cluster formation model predicts 
slow star formation.

If the star formation rate per free-fall time can be estimated from the
observations, then 
the number of protostars formed in a free-fall time, $N$, is given by 
\begin{eqnarray}
N &\simeq& M_{\rm cl} {\rm SFR}_{\rm ff}^{\rm obs} / M_{*}  \nonumber \\
 &=& 10 
\left({{\rm SFR}_{\rm ff}^{\rm obs} \over 0.01}\right)
\left({M_{\rm cl} \over 500 \ M_\odot}\right) 
\left({M_* \over 0.5 \ M_\odot}\right)^{-1}   \ ,
\label{eq:number of stars}
\end{eqnarray}
where $M_*$ is the mean mass of a protostar 
assuming the standard stellar IMF 
and here we adopt $M_{*} \simeq 0.5 M_\odot$, 
following \citet{chabrier05}.
Then, the average number of protostars observed at any given time 
can be estimated as 
\begin{equation}
 N_{\rm obs} \simeq N {t_{\rm life} \over t_{\rm ff}} \  ,
\end{equation}
where $t_{\rm life}$ is the typical lifetime of protostars.
Assuming that the protostars are the Class 0/I objects, 
$t_{\rm life}$ is estimated to be 
$t_{\rm life} \sim 0.4$ Myr \citep{evans09}.
From the observed number of Class 0/I objects, we can derive
the star formation rate per free-fall time from 
\begin{equation}
{\rm SFR}_{\rm ff}^{\rm obs} \simeq 0.01 \left({N_{\rm obs} \over 20}\right)
\left({M_{\rm cl} \over 500 \ M_\odot}\right)^{-3/2}
\left({M_* \over 0.5 \ M_\odot}\right)
\left({t_{\rm life} \over 0.4 \ {\rm Myr}}\right)^{-1}
\left({R_{\rm cl} \over 0.5 \ {\rm pc}}\right)^{3/2} \ .
\label{eq:sfr_obs}
\end{equation}

\subsection{Dynamical Impact of Protostellar Outflow Feedback}

To assess how the outflow feedback influences the clump dynamics,
we apply the  virial analysis to the cluster-forming clump.
The virial equation of a spherical clump is written as 
\begin{equation}
 {1 \over 2}\frac{\partial  ^2 I}{\partial t^2} = 2 U + W , 
\end{equation}
where $I$ is the moment of inertia, $U$ is the clump kinetic energy,
and $W$ is the clump gravitational energy.
The term $U$ and $W$ are given, respectively, by 
\begin{eqnarray}
  U &=& {3M_{\rm cl} \sigma_{\rm 1D}^2 \over 2} \\
 &=& 750 M_\odot \ {\rm km^2 \ s}^{-2} 
\left(M_{\rm cl} \over 500 \ M_\odot \right) \left(\sigma_{\rm 1D}
      \over {\rm km \ s}^{-1}\right)^2  \  ,
\end{eqnarray}
and
\begin{eqnarray}
 W &=& -{3 \over 5}af_B{GM^2 \over R} \\
 &\simeq& -1075 M_\odot {\rm km^2 \ s}^{-2}
\left(a \over 5/3 \right)\left(f_B \over 0.5\right)
\left(M_{\rm cl} \over 500 \ M_\odot\right)^2 \left(R_{\rm cl} \over 0.5
                                              \ {\rm pc} \right)^{-1}  \ .
\end{eqnarray}
In the outflow-regulated cluster formation model, the clump is in
quasi-virial equilibrium, i.e., $\alpha_{\rm vir} \sim 1$, where
$\alpha_{\rm vir}$ is the virial parameter defined as
\begin{eqnarray}
 \alpha_{\rm vir} &\equiv& - {2U \over W} = {5 \sigma_{\rm 1D}^2 R_{\rm cl} 
\over a f_B GM_{\rm  cl}}   \\
&\simeq& 1.4
\left({a \over 5/3}\right)^{-1} \left(f_B \over 0.5\right)^{-1}
\left({\sigma_{\rm 1D} \over {\rm km \ s}^{-1}}\right)^2
\left({R_{\rm cl} \over 0.5 \  {\rm pc}}\right)
\left({M_{\rm cl} \over 500 \ M_\odot}\right)^{-1} \ .
\end{eqnarray}

If the total outflow kinetic energy, $E_{\rm out}$, is significantly smaller than
the clump kinetic and gravitational energies, $U$ and $-$W, then
the outflow energy injection contributes little to the clump dynamical state.
To evaluate the impact of outflow feedback on the clump dynamics, 
we introduce the following non-dimensional quantity,
\begin{equation}
 \eta_{\rm out} \equiv -{2E_{\rm out} \over W}    \ .
\end{equation}
If $\eta_{\rm out}$ is as small as $\sim$ 0.1, then the outflow feedback 
is expected to play only a minor role in the clump dynamics.
On the other hand, if $\eta_{\rm out}$ is  larger than $\sim 1$,
the role of outflow feedback would be much more significant:  
the clump material is expected to be dispersed away from the parent
clump, and subsequent star formation is suppressed.

Here we consider only protostellar outflow feedback as the potential clump
disruption mechanism. This limits the clump mass to less than about 3000 M$_\odot$, above which UV radiation from O stars 
is likely to dominate the clump destruction assuming the standard
stellar IMF and star formation efficiencies \citep{matzner00,matzner02}.

\section{Confronting Model with Molecular Outflow Surveys}
\label{sec:obs}

Recently, several extensive molecular outflow surveys using the
$^{12}$CO lines have been carried out toward 
nearby cluster-forming regions. 
Here, we compare some characteristics of the outflow-regulated 
cluster formation model with the survey results toward 8 
nearby cluster-forming clumps, (1) B59, (2) L1551, (3) L1641N, 
(4) Serpens Main Cloud, (5) Serpens South, (6) $\rho$ Oph, 
(7) IC 348, and (8) NGC 1333. 
The masses of the clumps range from 
a few $\times 10 M_\odot$ to $10^3 M_\odot$.
All the cluster-forming clumps have distances smaller than about 400 pc,
so that the molecular outflows can be identified in reasonable 
spatial resolutions even with single-dish telescopes.
Since all the clumps contain no massive stars that would emit strong UV 
radiation, the outflow feedback is expected to be the leading stellar 
feedback mechanism in these regions.
 
It is worth noting that \citet{arce11} carried out detailed analysis of
large-scale $^{13}$CO ($J=1-0$) mapping data toward the Perseus molecular cloud and 
found a number of parsec-scale expanding bubbles that are presumably
driven by stellar winds from intermediate-mass protostars.
These bubbles are also expected to contribute to turbulence driving 
in the clouds. Similar bubbles are also found in
other regions like the Orion A molecular cloud \citep{heyer92,nakamura12}

\begin{figure}[h]
\epsscale{0.8}
\plotone{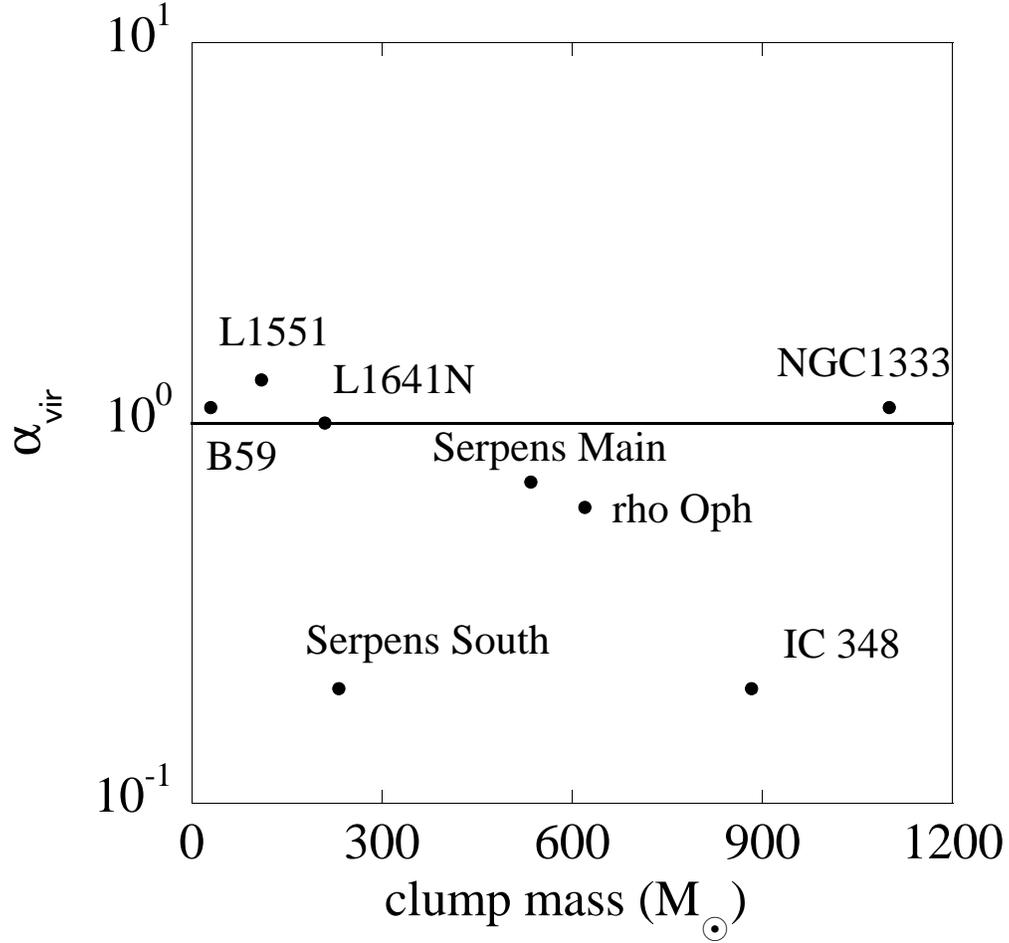}
\caption{Virial parameters of nearby cluster-forming clumps as a function
 of clump mass. To estimate the virial parameters, we neglect the
 possible magnetic support. 
In this sense, the derived values show lower limits.
We note that the physical quantities such as clump masses,
 sizes, and velocity dispersions are estimated with the
 $^{13}$CO ($J=1-0$) data except for Serpens Main and Serpens South.
For Serpens Main and Serpens South, the C$^{18}$O ($J=1-0$) and
N$_2$H$^+$ and Herschel data are used, respectively.
See the text for detail.
}  
\label{fig:virial}
\end{figure}

We present a brief summary of the 
molecular outflow surveys toward these 8 clumps in Table
\ref{tab:obssummary}, where the target name, observed $^{12}$CO transition,
telescope, receiver, observed period, velocity resolution ($\Delta V$), 
effective angular resolution ($\theta_{\rm eff}$), and rms noise level 
($\Delta T_{\rm mb}$) are listed.
Physical parameters of these clumps derived from the observations are 
also summarized in Table \ref{tab:target}, where the distance assumed, mass, radius,
mean surface density, 1D velocity dispersion, clump kinetic energy, gravitational energy,
virial parameter, molecular line used, and reference are presented.
The masses and radii are rescaled from the values presented in the literature
by taking into account the different distances assumed.
The virial parameters obtained from the observations are shown as a
function of clump mass in Figure \ref{fig:virial}.
For B59, we only consider the central round clump and do not take into
account the north-east and U-shape ridges (see Figure 1 of \citet{duarte12}).
We note that all clumps except Serpens Main and Serpens South were 
observed with the same molecular line tracer $^{13}$CO, whereas 
Serpens Main and Serpens South were observed with different molecular 
line tracers, C$^{18}$O and N$_2$H$^+$, respectively. 
In general, the velocity dispersions derived from these high density
tracers are smaller than those derived with $^{13}$CO.
But, its effect is expected to be minor for the estimation 
of the virial parameters because the clumps are centrally-condensed, 
and thus the virial parameters 
are not so sensitive to the density. In fact, for a clump with 
$\rho \propto r^{-2}$ and constant velocity dispersion, 
the virial parameter is a constant independent of radius. 
The other quantities such as $dP_{\rm turb}/dt$, virial velocity,
$dP_{\rm out}/dt$ predicted by the model are also constant for all
radii
when the density distribution follows $\rho \propto r^{-2}$.
It is worth noting that for the Serpens South clump, CO appears to be highly
depleted (Nishitani et al. 2014, in prep.) 
and $^{12}$CO and $^{13}$CO are significantly self-absorbed.
Even the C$^{18}$O emission does not follow the dense clump well. 
So, we adopt the mass estimated from the Spectral Energy Distribution
(SED) fitting of the {\it Herschel} data
\citep{tanaka13}, for which the column density is summed up in the area 
enclosed by a contour line of $5\times 10^{22}$ cm$^{-2}$.
For all clumps, the virial parameters are also estimated by omitting 
the effect of magnetic field, i.e., $f_B = 1$ is assumed.

Table \ref{tab:target} shows that the virial parameters of the
clumps are close to unity for all the clumps except for the Serpens South and 
$\rho$ Oph clumps.
This suggests that almost all the clumps are not far from the 
virial equilibrium. For Serpens South and $\rho$ Oph, the virial
parameters are found to be very small, $\alpha_{\rm vir} \sim 0.2$. 
For the Serpens South clump, \citet{tanaka13}
found that the infall motions are too slow compared to the free-fall
velocity, implying that the magnetic support may play a role 
in the clump support \citep[see also][]{sugitani11} and 
therefore we speculate that the ``effective'' 
virial parameter including the effect of magnetic support is close to unity.
The $\rho$ Oph clump also has a relatively small velocity dispersion,
which is much smaller than the free-fall velocity. However,
the magnetic field appears not to be spatially well-ordered
\citep{tamura11}, suggesting that the magnetic field may not be as 
important for the clump support as in Serpens South.
It remains unclear why $\rho$ Oph has a small virial parameter and 
appears to be relatively quiescent.  
In fact, only one Class 0 object, VLA1623, is found in the clump.
The $\rho$ Oph clump has very high visual extinction \citep{enoch07}, which may suggest
the presence of foreground molecular gas, or 
the $\rho$ Oph clump may be elongated 
along the line-of-sight, which would lead  to an under-estimate of the 
virial parameter and an over-estimate of the angle dispersion of the 
near IR polarization vectors.
The deficiency of Class 0 objects may be due to high extinction or, 
the star formation activity may be temporarily inactive \citep{enoch09}. 
Either way, the $\rho$ Oph clump is somewhat different from the other
cluster-forming clumps in our sample.

In the following, we will use the observational data to address two 
specific questions that lie at the heart of the outflow-regulated 
cluster formation model: (1) Does the outflow feedback has 
enough momentum to supply the dissipated turbulent motions? 
(2) Is the star formation in the surveyed region fast or slow? In addition, 
we will try to determine whether the outflow feedback has enough kinetic 
energy to unbind the parent clump, which is required in the competing 
model of rapid cluster formation.

\subsection{Outflow-Generated Turbulence}

\begin{figure}[h]
\epsscale{0.8}
\plotone{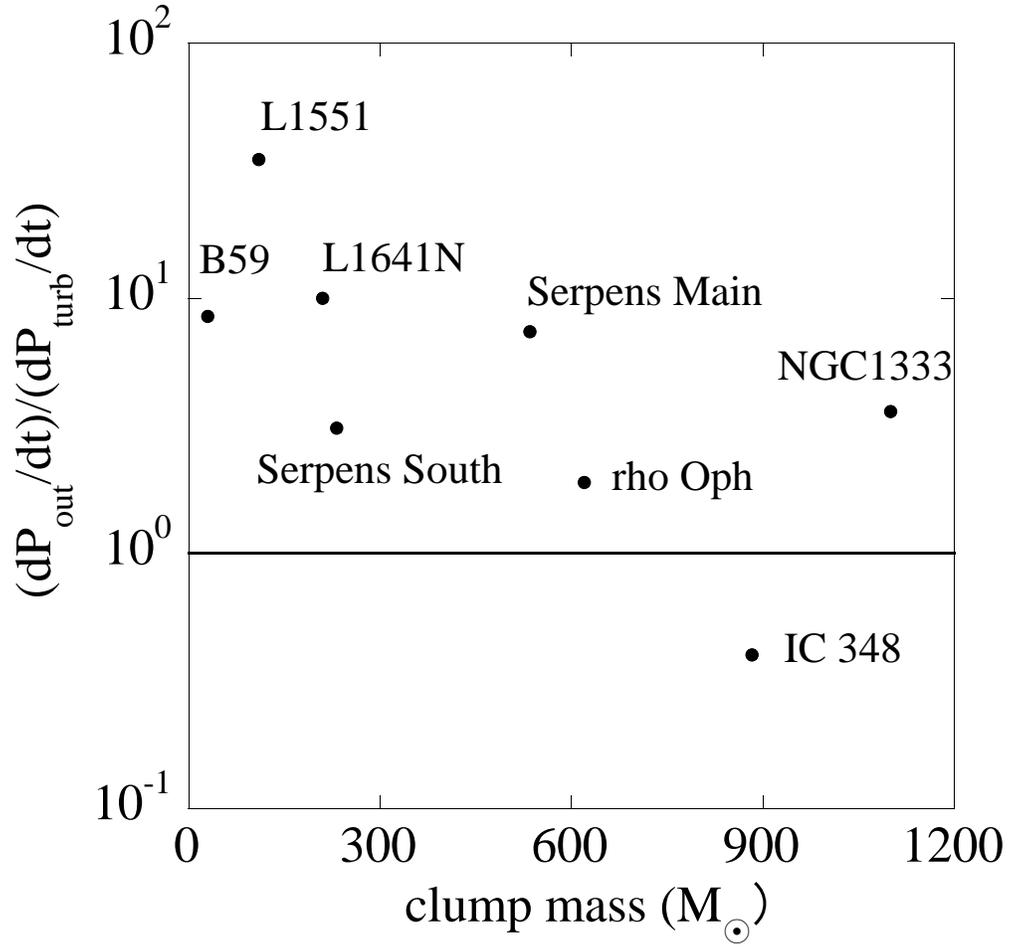}
\caption{Ratios of the momentum injection to turbulence dissipation
 rates of nearby cluster-forming clumps as a function
 of clump mass. 
}  
\label{fig:dp/dt}
\end{figure}

In the outflow-regulated cluster formation model, the turbulence
momentum dissipation rate should balance the outflow momentum
injection rate.  In Table \ref{tab:outflow lobe}, we present
the turbulence momentum dissipation rate $dP_{\rm turb}/dt$, 
the outflow momentum injection rate $dP_{\rm out}/dt$, their ratio 
$(dP_{\rm out}/dt)/(dP_{\rm turb}/dt)$, the outflow momentum, the outflow kinetic energy 
$E_{\rm out}$, and the ratio $\eta_{\rm out}$ for all 8 cluster-forming clumps.
The outflow parameters are derived from the quantities presented in the
references shown in 
Table \ref{tab:obssummary}
after applying corrections to account for different distances adopted 
and different assumptions.
To estimate these quantities, we assume the following:
(1) the inclination angles of the outflow axes are around $\xi \simeq
57.3 ^\circ$, (2) the outflow material is optically-thin, and
(3) the typical dynamical time of the outflows is $3\times 10^4$ year.
The second assumption leads to an under-estimation of 
the outflow momentum injection rates and the outflow kinetic 
energies by a factor of a few or more. Also, the outflow momenta and energies are 
underestimated at least by a factor of a few because the 
low-velocity components are omitted due to the difficulty in separating 
such outflow components from 
the ambient clump material \citep{bally99,arce10,offner11}.
For example, for $\rho$ Oph, the emission whose LSR velocity is in the
range from 1 km s$^{-1}$ to 6.5 km s$^{-1}$ is not taken into account 
in estimating the physical parameters of the outflows, although this velocity width of 
5.5 km s$^{-1}$ is much larger than the FWHM velocity width of the clump
(1.5 km s$^{-1}$). For the Serpens Main Cloud and Serpens South, the emissions with 
6 km s$^{-1}$ $ \le V_{\rm LSR} \le 10$ km s$^{-1}$ 
and 4 km s$^{-1}$ $ \le V_{\rm LSR} \le 11$ km s$^{-1}$ are omitted,
respectively, and the velocity widths of 4 km s$^{-1}$ and 7 km s$^{-1}$ are about twice 
and 5 times larger than the FWHM velocity widths of 2 km s$^{-1}$ and
1.2 km s$^{-1}$, respectively.
In total, the outflow masses, momenta, and kinetic energies derived from
the observations are likely to be underestimated by an order of magnitude.
This underestimation of the outflow physical quantities may be compensated
by the factor $f_{\rm out}$, which is the fraction of the molecular
outflow momentum that is converted into the clump turbulent momentum, and is
expected to be around a few $\times 10^{-1}$.
Therefore, we assume that the momentum injection rates derived from the
observations (presented in Table \ref{tab:outflow lobe}) are comparable
to the outflow momentum injection rate.
We present the ratio $(dP_{\rm out}/dt)/(dP_{\rm turb}/dt)$ as a
function of clump mass in Figure \ref{fig:dp/dt}.

According to Table \ref{tab:outflow lobe} and Figure \ref{fig:dp/dt}, 
for all clumps except $\rho$ Oph, the outflow momentum injection
rate is comparable to or larger than the turbulence dissipation rate. 
Therefore, we conclude that the outflows can maintain supersonic turbulence 
in the cluster-forming clumps. 
For the three least massive clumps, B59, L1551, and L1641N, the ratios 
between $dP_{\rm out}/dt$ and $dP_{\rm turb}/dt$ tend to be larger. 
This is presumably due to the fact that these clumps are 
nearest and the numbers of outflow lobes and protostars are smaller, and thus 
it is easier to distinguish between the outflow components and 
ambient clump gas.

We note that in previous studies, the energy dissipation and injection
rates are compared to assess whether the outflow feedback can maintain
the turbulent motions in the cluster-forming clumps. The main
conclusion is that the energy injection rate due to the outflow
feedback is generally larger than the energy dissipation rate, and thus 
the outflow feedback has enough energy to maintain the turbulent motions
\citep[e.g.,][]{hatchell07,stanke07,swift08,maury09,arce10,curtis10,nakamura11a,nakamura11b}. 
However, the outflow feedback is a momentum-driven feedback because 
radiative energy loss is efficient in the clouds and clumps \citep{fall10}.
Thus, our approach of using the momentum dissipation and injection rates 
is likely to be more appropriate.

\subsection{Dynamical Impact of Outflow Feedback}

As shown in Table \ref{tab:target}, almost all clumps appear 
close to virial equilibrium. The outflow feedback should provide additional force in 
the clump material.
If the outflows have enough energies to disperse the surrounding gas,
then the outflow feedback can quench further star formation.
Here, we measure the dynamical effect of the outflow feedback 
in the clump destruction, using the non-dimensional parameter
$\eta_{\rm out}$,
the ratio between 2$E_{\rm out}$ and $-W$.

\begin{figure}[h]
\epsscale{0.8}
\plotone{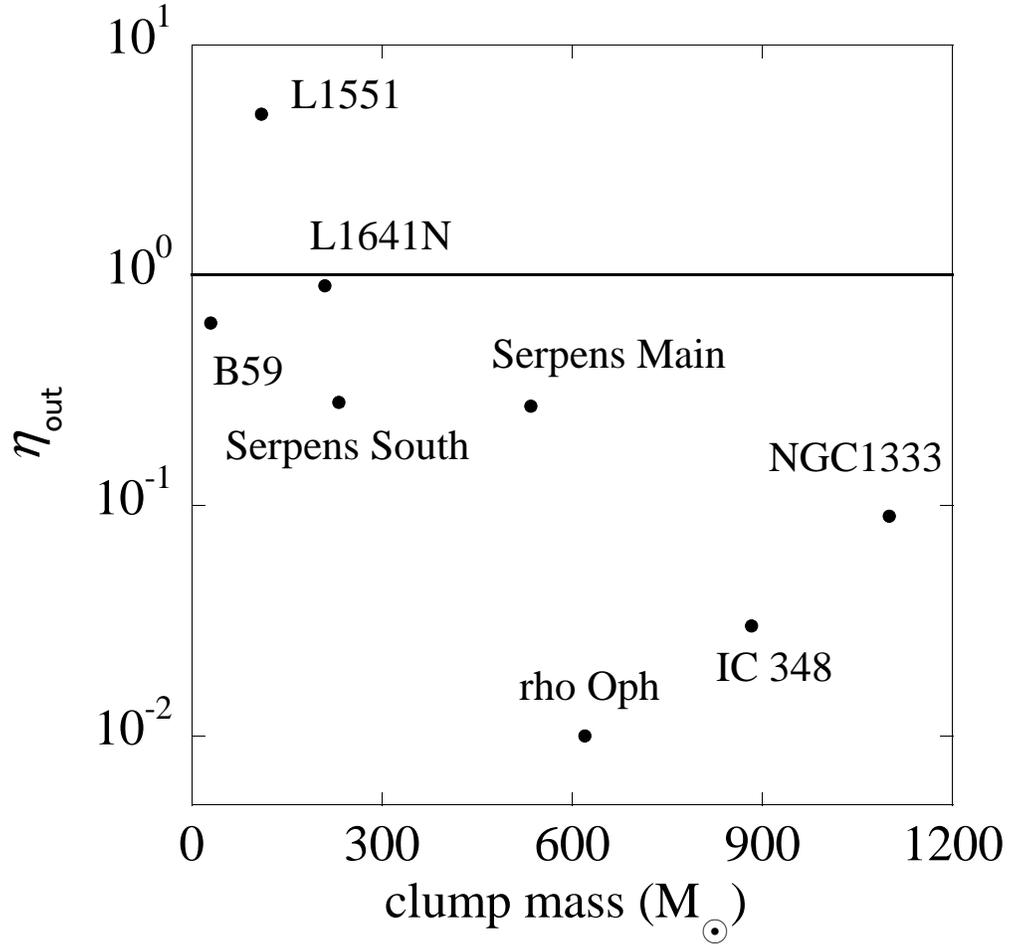}
\caption{Ratios of twice the outflow kinetic energy $2E_{\rm out}$ to the gravitational binding energy of the clump $-W$, $\eta_{\rm out}$, of nearby cluster-forming clumps as a function
 of clump mass. 
}  
\label{fig:eta}
\end{figure}

In the last column of Table \ref{tab:outflow lobe} 
and Figure \ref{fig:eta}, 
we present 
the values $\eta_{\rm out}$ derived from the observations.
We note that presumably only a fraction of $E_{\rm out}$ contributes to
the dispersal of the clump material because a significant fraction of 
the outflow kinetic energy escapes out of the clump once the outflow 
breaks out.
This fraction is expected to be larger for less massive, smaller clumps.
However, this effect may be compensated by the fact that the outflow kinetic
energies derived from the observations are likely to be underestimated
by an order of magnitude.
Therefore, we use the values of $\eta_{\rm out}$ presented in Table \ref{tab:outflow lobe}
to assess the dynamical impact of the outflow feedback.

For the three least massive clumps, B59, L1551, and L1641N, 
the values of $\eta_{\rm out}$
are large, indicating that the outflow feedback has potential
to impact the clump structure and dynamics significantly.
For L1641N, there is evidence that 
the stellar feedback may have dispersed the clump material significantly
\citep{reipurth98,nakamura12}.
For the intermediate mass clumps such as the Serpens Main Cloud and
Serpens South, the outflow kinetic energy may partly influence the 
clump dynamical evolution. 
In contrast, for massive clumps, the outflow feedback appears to
play a minor role in the global clump dynamics.
In other words, it is likely difficult to destroy 
the whole clumps  only by the current outflow activity.
This suggests that whether the outflow feedback can destroy
the cluster-forming clumps or not may depend on the clump
mass.  For massive clumps, the outflow feedback appears unable 
to disperse the clump material significantly and thus the star 
formation may proceed a relatively long time.

\subsection{Star Formation Rate Per Free-Fall Time}

\begin{figure}[h]
\epsscale{0.8}
\plotone{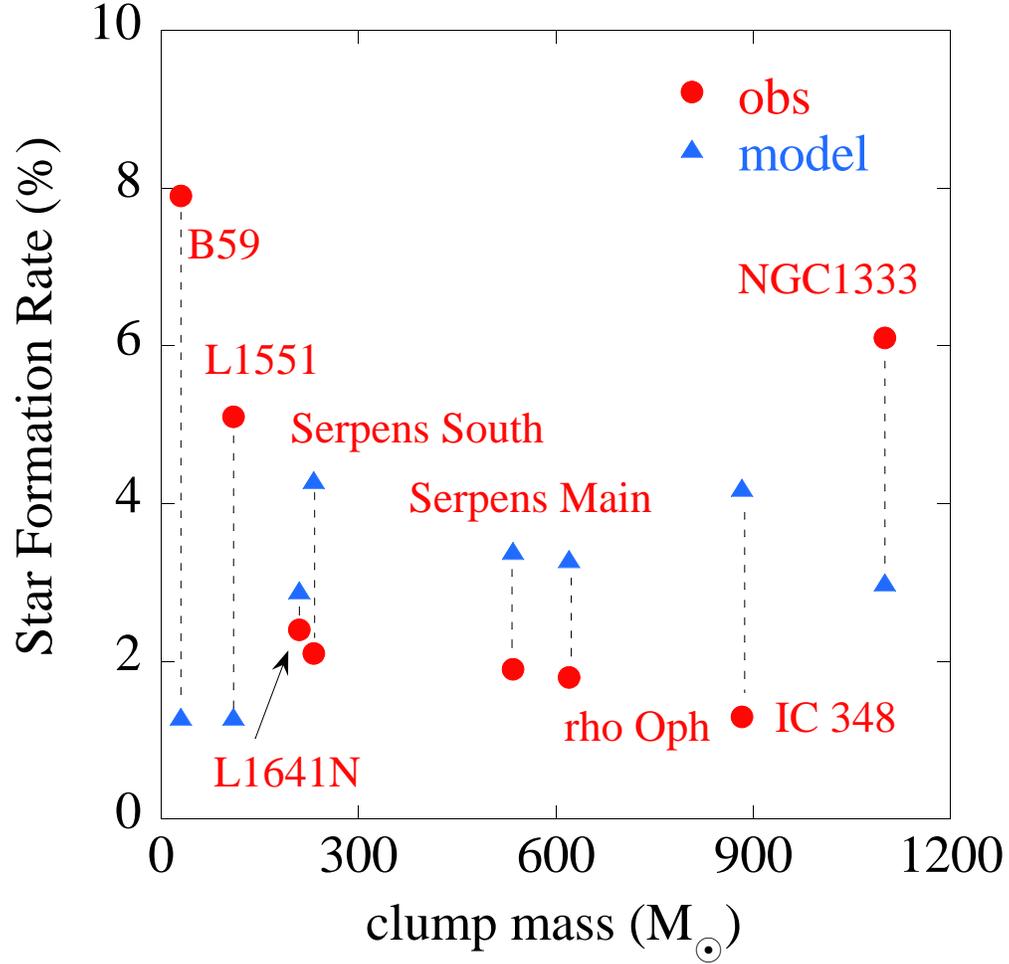}
\caption{Star formation rates per free-fall time of nearby
 cluster-forming clumps. The red circles and blue triangles indicate
SFR$_{\rm ff}$ derived from observations and SFR$_{\rm ff}$ predicted by
 the outflow-regulated cluster formation model, respectively.
}  
\label{fig:sfr}
\end{figure}

Table \ref{tab:sfr} summarizes the number of protostars (Class 0 and I objects)
observed in the individual clumps, and the star formation rates
per free-fall time derived from the observations and Equation
(\ref{eq:sfr_ff}). 
The star formation rates per free-fall time derived from the observations
and predicted by the model are presented in Figure \ref{fig:sfr}.
For B59, L1551, and L1641N, we adopt the results of \citet{brooke07},
\citet{stojimirovic06}, and \citet{megeath12}, respectively.
For the Serpens Main Cloud and $\rho$ Oph, the numbers of protostars are 
calculated using the results of the {\it Spitzer} observations \citep{evans09}. 
For Serpens South, we count the protostars located within the
circle indicated in Figure 1 of 
\citet{gutermuth08}. We also add the Class 0 sources
identified within the circle by \citet{bontemps10}. 
For IC 348 and NGC 1333, we adopt the numbers shown in \citet{arce10}.
Here, we adopt the typical lifetime for Class I objects of 
0.4 Myr on the basis of the results of the {\it Spitzer} Gould Belt
Survey \citep{evans09}, although the lifetime may depend on the
interstellar environments somewhat.
For all clumps, the observed SFR$_{\rm ff}^{\rm obs}$ and 
SFR$_{\rm ff}$ predicted by the outflow-regulated cluster formation
model stays as low as a few percent.
Taking into account that the quantities have uncertainty at least by a
factor of a few, we conclude that they are consistent with each other, supporting 
the slow cluster formation model.
If SFR$_{\rm ff} = 10 \sim$ a few $\times 10$ \% as suggested by
the rapid cluster formation model, the lifetime of the Class I objects 
should be of order of $10^4$ yr, which is too short \citep{evans09}.

\section{Conclusion}
\label{sec:conclusion}

In the present paper, we constructed an analytic model of the
outflow-regulated cluster formation scenario and confronted 
some of the model predictions with recent outflow surveys toward 8 nearby
cluster-forming clumps: B59, L1551, L1641N, Serpens Main Cloud, Serpens
South, IC 348, and NGC 1333. We found that the observational results support 
the outflow-regulated cluster formation model in general. 
The main conclusions are summarized below.

\begin{itemize}

\item[1]
We constructed an analytic model of the outflow-regulated cluster
       formation, in which we assumed that the turbulence dissipation
       rate is balanced by the outflow momentum injection
       rate in a cluster-forming clump
that is in virial equilibrium.  In this model, the star formation rate
        per free-fall time is predicted to be a few percent.

 \item[2] Most of the surveyed cluster-forming clumps have virial parameters 
close to unity, indicating that the internal 
turbulent motions play an important role in the clump support, and 
that the clumps are close to virial equilibrium in general. The exceptions are 
Serpens South and $\rho$ Oph, where the virial parameters are
estimated to be as small as $\sim$ 0.2.
In Serpens South, \citet{sugitani11} revealed the existence of 
globally-ordered magnetic field that appears to be roughly
perpendicular to the main filament, indicating that the magnetic support
is important \citep[see also][]{tanaka13}. In contrast, 
for $\rho$ Oph, no globally-ordered magnetic field has been 
observed \citep{tamura11}. However, 
the clump does not appear to be globally collapsing at the free-fall 
rate despite its slow internal turbulent motions. It remains unclear why $\rho$ Oph 
appears relatively quiescent. It might be elongated along the line-of-sight, so that
the virial parameter is underestimated.

\item[3] For most of the clumps, the outflow momentum injection rate
is comparable to or larger than the turbulence momentum dissipation
rate. We note that the outflow momenta are underestimated in this paper because 
the outflow gas is assumed to be optically-thin and the low-velocity outflow 
components are ignored. The actual outflow momentum injection rates should 
be larger by a factor of a few or more. Thus, we conclude that the outflow 
feedback can maintain supersonic turbulence in the surveyed nearby cluster-forming regions.

\item[4] 
However, the outflow kinetic energy is only a fraction of 
the clump gravitational energy except for the three least massive
         clumps, B59, L1551 and L1641N. 
Therefore, we conclude that the outflow feedback is not enough to 
disperse the whole clump at least for the intermediate-mass and massive clumps.

\item[5]
Using the numbers of Class 0/I objects, 
the star formation rates per free-fall time are estimated to be 
a few percent for all 8 clumps, which is consistent with
the outflow-regulated scenario of slow cluster formation.

\acknowledgements
FN is supported in part by a Grant-in-Aid for Scientific 
Research of Japan (A, 24244017), and ZYL by NASA NNH10AH30G and NNX14AB38G, and NSF AST-1313083. 

\end{itemize}

\clearpage

\begin{deluxetable}{lllllllll}
\tabletypesize{\scriptsize}
\tablecolumns{9}
\tablecaption{Summary o $^{12}$CO Outflow Surveys Toward 
Nearby Parsec-Scale Cluster-Forming Clumps}
\tablewidth{\columnwidth}
\tablehead{\colhead{Name} 
& \colhead{CO transition} 
& \colhead{telescope} 
& \colhead{receiver} 
& \colhead{period} 
& \colhead{$\Delta V$} 
& \colhead{$\Delta \theta_{\rm eff}$} 
& \colhead{$\Delta T_{\rm mb}$} 
& \colhead{Reference$^e$} \\
\colhead{} 
& \colhead{} 
& \colhead{} 
& \colhead{} 
& \colhead{} 
& \colhead{(km s$^{-1}$)} 
& \colhead{($''$)} 
& \colhead{(K)} 
& \colhead{} 
}
\startdata
B59 & $3-2$  & JCMT  & HARP & 2010.5 $-$ 2010.6  
 & 0.5 & 20 & 0.2 & 1 \\
L1551 & $1-0$  & FCRAO  & SEQUOIA & 2001 $-$ 2002  
 & 0.25 & 50 & 0.2 & 2 \\
L1641N & $1-0$  & NRO  & BEARS & 2009.12$-$2010.1  
 & 0.5 & 21 & 1.2 & 3 \\
Serpens Main & $3-2$ & JCMT & HARP & 2007.4, 2007.7  & 1.0  & 20 & 0.16  & 4 \\
Serpens South & $3-2$ & ASTE & MAC 345  & 2010.8 
 & 0.5 & 24 & 0.19 & 5 \\
$\rho$ Oph & $1-0$ & NRO  & BEARS & 2009.12$-$2010.5 
 & 0.4  & 30 & 1.0 & 6 \\
IC 348 &$1-0$  & FCRAO  & SEQUOIA & 2002$-$2005 
&  0.07 & 50 & 0.5 & 7 \\
NGC1333 &$1-0$  & FCRAO  & SEQUOIA & 2002$-$2005 
&  0.07 & 50 & 0.5 & 7
\enddata
\tablenotetext{a}{1. \citet{duarte12}; 2. \citet{stojimirovic06}; 
3. \citet{nakamura12}; 
4. \citet{graves10}; 5. \citet{nakamura11b}; 
6. \citet{nakamura11a}; 7. \citet{arce10}}
\tablecomments{See Section \ref{sec:obs} in detail.}
\label{tab:obssummary}
\end{deluxetable}

\begin{deluxetable}{lllllllllll}
\tabletypesize{\scriptsize}
\rotate
\tablecolumns{9}
\tablecaption{Physical Parameters of Nearby Parsec-Scale Cluster-Forming Clumps}
\tablewidth{\columnwidth}
\tablehead{\colhead{Name} 
& \colhead{Distance} 
& \colhead{Mass} 
& \colhead{Radius} 
& \colhead{$\Sigma$} 
& \colhead{$\sigma_{\rm 1D}$} 
& \colhead{$U$} 
& \colhead{$W$} 
& \colhead{$\alpha_{\rm vir}$} 
& \colhead{molecular line} 
& \colhead{Reference} \\
\colhead{} 
& \colhead{(pc)} 
& \colhead{($M_\odot$)} 
& \colhead{(pc)} 
& \colhead{(g cm$^{-2}$)} 
& \colhead{(km s$^{-1}$)} 
& \colhead{($M_\odot$ km$^2$s$^{-2}$)} 
& \colhead{($M_\odot$ km$^2$s$^{-2}$)} 
& \colhead{} 
& \colhead{} 
& \colhead{} 
}
\startdata
B59 & 130 & 30  & 0.3 & 0.17 & 0.4 & 7 & 13 & 1.1  &
 $^{13}$CO ($1-0$) & 1 \\
L1551 & 140 & 110  & 1 & 0.007 & 0.45 & 33 & 52 & 1.3  & $^{13}$CO ($1-0$) & 2 \\
L1641N & 400 & 210 & 0.55 & 0.045 & 0.74 & 214 & 581  & 1.0 
& $^{13}$CO ($1-0$) & 3 \\
Serpens Main & 415 & 535 & 0.73 & 0.065 & 0.85 & 580 & 1686 & 0.7 & C$^{18}$O ($1-0$)  &
 4,5 \\
Serpens South & 415  & 232  & 0.2 & 0.38 & 0.53 & 98 & 1157  & 0.2  
& {\it Herschel}, N$_2$H$^+$ ($1-0$)& 6 \\
$\rho$ Oph & 125 & 883 & 0.8 & 0.090 & 0.64  & 543 &4191 & 0.2 & $^{13}$CO ($1-0$) & 7 \\
IC 348 & 250 & 620 & 0.9 & 0.050 & 0.76 & 753 & 1837  & 0.6 & $^{13}$CO ($1-0$) & 8 \\
NGC 1333 & 250 & 1100 & 2.0 & 0.018 & 0.93  & 1427& 2602 & 1.1  & $^{13}$CO ($1-0$)  & 8   
\enddata
\tablenotetext{a}{1. \citet{duarte12}; 2. \citet{stojimirovic06}; 
3. \citet{reipurth98};  4. \citet{olmi02}; 5. \citet{sugitani10}; 
6. \citet{tanaka13}; 7. \citet{nakamura11a}; 
8. \citet{arce10}}
\label{tab:target}
\end{deluxetable}

\begin{deluxetable}{lllllll}
\tabletypesize{\scriptsize}
\tablecolumns{7}
\tablecaption{Observations of Nearby Parsec-Scale Cluster-Forming Clumps}
\tablewidth{\columnwidth}
\tablehead{\colhead{Name} 
& \colhead{$dP_{\rm turb}/dt$} 
& \colhead{$dP_{\rm out}/dt ^a$} 
& \colhead{$(dP_{\rm out}/dt)/(dP_{\rm turb}/dt)$} 
& \colhead{${P}_{\rm out}^b$} 
& \colhead{${E}_{\rm out}^b$} 
& \colhead{${\eta}_{\rm out}$} \\
  \colhead{} 
& \colhead{($M_\odot$ km s$^{-1}$ yr$^{-1}$)} 
& \colhead{($M_\odot$ km s$^{-1}$ yr$^{-1}$)} 
& \colhead{} 
& \colhead{($M_\odot$ km s$^{-1}$)} 
& \colhead{($M_\odot$ km$^2$ s$^{-2}$)} 
& \colhead{}
}
\startdata
B59    & $1.0 \times 10^{-5}$ & $8.5\times 10^{-5}$ & 8.5  & 2.6 & 4  & 0.62   \\
L1551  & $1.8 \times 10^{-5}$ & $6.3\times 10^{-4}$ & 35 & 19 & 130 & 5.0   \\
L1641N & $1.3 \times 10^{-4}$ & $1.3\times 10^{-3}$ & 10  & 80  & 273 &  0.9  \\
Serpens Main & $3.4\times 10^{-4}$ & $2.5\times 10^{-3}$ & 7.4 & 75  & 445 & 0.27 \\
Serpens South & $2.1\times 10^{-4}$ & $6.5\times 10^{-4}$ & 3.1  & 19 & 165 & 0.28  \\
$\rho$ Oph & $2.9\times 10^{-4}$ & $1.2\times 10^{-4}$ & 0.4  & 3.6 & 61  &
 0.03 \\
IC 348   & $2.5\times 10^{-4}$ & $4.7\times 10^{-4}$ & 1.9  & 14  & 26  & 0.01 \\
NGC 1333 & $3.0\times 10^{-4}$ & $1.1\times 10^{-3}$ & 3.6  & 32  & 119 & 0.09   
\enddata
\tablenotetext{a}{The outflow momentum injection rates are 
highly underestimated. See Section \ref{sec:obs} in detail.
The dynamical time of $3\times 10^4$ yr is also adopted to
derive the outflow momentum injection rates.}
\tablenotetext{b}{The following two conditions are assumed: (1) the
 outflow gas is optically-thin, and (2) the outflow axes are randomly
 distributed in the plane-of-sky, and the mean inclination angle of 
$\xi = 57.3^\circ$ is applied for all the outflow components.}
\label{tab:outflow lobe}
\end{deluxetable}

\begin{deluxetable}{llll}
\tabletypesize{\scriptsize}
\tablecolumns{4}
\tablecaption{Star Formation in Nearby Parsec-Scale Cluster-Forming Clumps}
\tablewidth{\columnwidth}
\tablehead{\colhead{Name} 
& \colhead{$N_{\rm Class 0/I}$} 
& \colhead{SFR$_{\rm ff}^{\rm obs}$$^a$} 
& \colhead{SFR$_{\rm ff}$$^b$} \\
 \colhead{} 
& \colhead{} 
& \colhead{(\%)} 
& \colhead{(\%)} 
}
\startdata
B59 & 4 & 7.9 & 1.3  \\
L1551 & 3 & 5.1 & 1.3 \\
L1641N & 14 & 2.4 & 2.9 \\
Serpens Main & 14  & 1.9 & 3.4  \\
Serpens South & 42 & 2.1  & 4.3  \\
$\rho$ Oph & 23 & 1.3 & 4.2  \\
IC 348 & 16 & 1.8 & 3.3  \\
NGC 1333 & 40  & 6.1 & 3.0   
\enddata
\tablenotetext{a}{SFR$_{\rm ff}^{\rm obs}$ is derived
 from Equation (\ref{eq:sfr_obs}).}
\tablenotetext{b}{SFR$_{\rm ff}$ is derived
 from Equation (\ref{eq:sfr_ff}) with $f_B = 1$.}
\tablecomments{The lifetime of protostars is assumed to be 0.4 Myr for
 all the regions.}
\label{tab:sfr}
\end{deluxetable}

\clearpage

\end{document}